\documentclass[journal=jpccck,manuscript=article]{achemso}
\usepackage{amsfonts,amssymb,xspace,color}
\usepackage{amsmath}
\usepackage{setspace}
\usepackage{graphicx}
\usepackage{subfigure}
\usepackage{float}
\usepackage{pifont}

\usepackage{GlavDesign}

\newcommand{\scaleprofile}{0.7}

\newcommand{\ho}{H$_2$O }
\newcommand{\co}{CO$_2$ }
\newcommand{\ch}{CH$_4$ }
\newcommand{\coo}{CO$_2$}
\newcommand{\chh}{CH$_4$}
\newcommand{\coeq}{{\mathrm{CO_2}}}
\newcommand{\cheq}{{\mathrm{CH_4}}}
\newcommand{\waeq}{{\mathrm{H_2O}}}

\newcommand{\myref}[1]{\plainref{#1}}
\renewcommand{\eqr}[1]{Eq. (\myref{#1})}
\renewcommand{\secr}[1]{Section [\myref{#1}]}
\renewcommand{\figr}[1]{Figure [\myref{#1}]}
\newcommand{\figsr}[2]{Figures [\myref{#1}] and [\myref{#2}]}
\renewcommand{\tblr}[1]{Table [\myref{#1}]}
\renewcommand{\eqref}[1]{(\myref{#1})}

\title{Gibbs and Helmholtz energies of formation of sI clathrate hydrates from \co, \ch and water.}

\author{K.~S.~Glavatskiy\tsup{1,2}}
\email{kirill.glavatskiy@rmit.edu.au} %
\author{T.~J.~H.~Vlugt\tsup{2}}
\author{S.~Kjelstrup\tsup{1,2}}
\affiliation {
\tsup{1}Department of Chemistry, Norwegian University of Science and Technology, NO 7491 Trondheim, Norway.\\
\tsup{2}Department of Process and Energy, Delft University of Technology, Leeghwaterstraat 44, 2628 CA Delft, The Netherlands. }
\date\today

\SectionNumbersOn

\begin{document}

\begin{abstract}
We determine  thermodynamic stability conditions in terms of Helmholtz and Gibbs energies for sI clathrate hydrates with \ch and \co at 278 K.  Helmholtz energies are relevant for processing from porous rocks (constant volume), while Gibbs energies are relevant for processing from layers on the ocean floor (constant pressure). We define three steps leading to hydrate formation, and find Helmholtz energy differences from molecular simulations for two of them using grand-canonical Monte Carlo simulations  at constant temperature and volume; while the third step was calculated from literature data. The Gibbs energy change for the same steps are also determined.   From the variations in the total Helmholtz and Gibbs energies we suggest thermodynamic paths for exchange of \ch by \co in the isothermal hydrate, for constant volume or pressure, respectively.  We show how these paths for the mixed hydrate can be understood from single-component occupancy isotherms, where \co, but not \ch, can distinguish between large and small cages. The strong preference for \ch for a range of compositions can be explained by these. 
\end{abstract}

\maketitle



%
\section{Introduction}\label{sec/Introduction}

Over the last years, it has become increasingly clear that a substantial part of the world's gas resources is connected to sI clathrate hydrates, or more simply, hydrates.  These occur either in the form of layers, say on the ocean floor, but also inside porous rocks \cite{Baldwin2009, Clarke1999}.  In the first case, the pressure will be constant, while in the last case, the hydrate is confined in a constant volume. Efficient processing depends on stability information for both conditions. Stability criteria for hydrates under various conditions are therefore of central interest, giving a motivation for the present work, started recently \cite{Glavatskiy2012}. 

The formation of a hydrate containing methane and carbon dioxide from its fluid components can be we written as:
\begin{equation}\label{eq/Intro/01}
\left[N_{\cheq} + N_{\coeq} + N_{\waeq}\right]_{\mathrm{fluid}} \rightleftharpoons \left[(N_{\cheq} + N_{\coeq}) \cdot N_{\waeq}\right]_{\mathrm{hydrate}}
\end{equation}
where $N_{\cheq}$, $N_{\coeq}$ and $N_{\waeq}$ are the number of methane, carbon dioxide and water molecules, respectively. The composition reflects a one-phase region in the phase diagram (REF).  
The gases will be referred to as guest molecules from now on. The subscripts $_\mathrm{fluid}$ and $_\mathrm{hydrate}$ indicates the phase which is formed by the corresponding number of \ch, \co and \ho molecules. In particular, subscript $_\mathrm{hydrate}$ means that the molecules form a sI hydrate, while subscript $_\mathrm{fluid}$ means that the same number of molecules form a disordered, essentially fluid phase (in the proper range of compositions).  
A negative or zero Helmholtz energy difference means that the hydrate is stable at constant volume, while a negative or zero Gibbs energy difference provides a stability criertion  for constant pressure. Both criteria apply at constant temperature. 

Knowledge of the stability of a hydrate with a mixture of gases, relative to pure gas hydrates, can furthermore be used to decide whether the idea of processing \ch, while storing \co, is feasible for conditions which are interesting from a practical perspective.  The aim of this work is to provide quantitative information on the possibility of this exchange, which has been studied experimentally \cite{Stevens}, and which is now on the point of being realized in Alaska \cite{alaskalink}.  

For computational purposes, it is convenient to divide the hydrate formation \eqref{eq/Intro/01} into three steps (REFS?):
\begin{subequations}\label{eq/Intro/02}
\begin{equation}\label{eq/Intro/02a}
\left[N_{\cheq} + N_{\coeq} + N_{\waeq}\right]_{\mathrm{fluid}} \rightleftharpoons \left[N_{\cheq} + N_{\coeq}\right]_{\mathrm{gas}} +\left[ N_{\waeq}\right]_{\mathrm{fluid}}\\
\end{equation}
\begin{equation}\label{eq/Intro/02b}
\left[N_{\waeq}\right]_{\mathrm{fluid}} \rightleftharpoons \left[ N_{\waeq}\right]_{\mathrm{hydrate}}\\
\end{equation}
\begin{equation}\label{eq/Intro/02c}
\left[N_{\cheq} + N_{\coeq}\right]_{\mathrm{gas}} +\left[ N_{\waeq}\right]_{\mathrm{hydrate}} \rightleftharpoons \left[(N_{\cheq} + N_{\coeq}) \cdot N_{\waeq}\right]_{\mathrm{hydrate}}\\
\end{equation}
\end{subequations}
The first step describes desorption of gas molecules from fluid water. In the second step, a pure hydrate network is formed from fluid water.  The empty hydrate is not stable, so this step is hypothetical. In the third step, the guest molecules of \co and/or \ch are encapsulated into the empty cages of the clathrate from the gas phase. 

The Helmholtz energy difference for step \eqref{eq/Intro/02c} was reported earlier \cite{Glavatskiy2012}.  In this paper we will add information on the two other steps, \eqr{eq/Intro/02a} and \eqr{eq/Intro/02b}, in order to obtain thermodynamic information on the total formation reaction \eqref{eq/Intro/01}. We shall obtain the Helmholtz energy of the first step by molecular simulations, and the second step by literature data. When added to previous results we obtain the stability criterion, $\Delta_r F$, for hydrate formation at constant volume and temperature (hydrate formation in porous rocks). Building on these data, we shall determine the Gibbs energy difference $\Delta_r G$ of the formation \eqref{eq/Intro/01}, relevant for gas processing from the ocean floor.  The details of the thermodynamic calculations are given in \secr{sec:Thermodynamics}. 

We introduce the following symbols for the energy differences: $\Delta_{f} F$, $\Delta_{w} F$, and $\Delta_{h} F$. Here $\Delta_{f} F$  denotes the difference between the Helmholtz energy of fluid water $\left[ N_{\waeq}\right]_{\mathrm{fluid}}$ with a gas phase containing of $N_{\cheq}$ and $N_{\coeq}$ molecules and the Helmholtz energy of the fluid with dissolved gas(es), $\left[N_{\cheq} + N_{\coeq} + N_{\waeq}\right]_{\mathrm{fluid}}$, \textit{i.e.} the Helmholtz energy of the step \eqref{eq/Intro/02a}. 

The symbol $\Delta_{w} F$ is the difference in the Helmholtz energy of the metastable hydrate structure, consisting of $N_{\waeq}$ water molecules, and the fluid water with the same number of water molecules, \textit{i.e.} the Helmholtz energy of the step \eqref{eq/Intro/02b}. Finally, $\Delta_{h} F$ is the difference between the Helmholtz energy of the hydrate with a mixed gas encaged, $\left[(N_{\cheq} + N_{\coeq}) \cdot N_{\waeq}\right]_{\mathrm{hydrate}}$ and the Helmholtz  energy of an empty hydrate $\left[ N_{\waeq}\right]_{\mathrm{hydrate}}$ with a gas phase containing $N_{\cheq}$ and $N_{\coeq}$ molecules, \textit{i.e.} the Helmholtz energy of step \eqref{eq/Intro/02c}.

The Helmholtz energy $\Delta_r F$ of reaction \eqref{eq/Intro/01} is equal to the sum
\begin{equation}\label{eq/Intro/03a}
\Delta_r F = \Delta_{f} F + \Delta_{w} F + \Delta_{h} F 
\end{equation}

The processes in \eqref{eq/Intro/02a} and in \eqref{eq/Intro/02c} can be compared to adsorption of small molecules in porous materials such as zeolites and metal organic frameworks, for which grand-canonical Monte Carlo (GCMC) simulations have been used extensively \cite{GarciaPerez2011, Castillo2009, Vlugt2008, GarciaPerez2007, Karavias1991}. The GCMC simulations, estalished for step \eqref{eq/Intro/02c}  \cite{Glavatskiy2012}, were identical to those used to study adsorption in zeolites. The procedures will be also used here to simulate $\Delta_{f} F$, while $\Delta_{w} F$ for the \eqref{eq/Intro/02b} will be taken from data in the literature. 
The values of the Helmholtz energy differences will next be used to find the corresponding Gibbs energy differences of the steps \eqref{eq/Intro/02a} - \eqref{eq/Intro/02c} using data from the literature.

The paper is organized as follows. In \secr{sec:Thermodynamics} we present the thermodynamic relations which give the Helmholtz energy and the Gibbs energy. The details of the simulations are specified in \secr{sec/Simulations}. In \secr{sec/Results} we provide the results and discuss their implications. In particular, we discuss the feasibility of a thermodynamic path which favors the exchange of \co with \ch in hydrates.

\section{Thermodynamics}\label{sec:Thermodynamics}

Consider first a hydrate phase in a closed volume,  $V$, and constant temperature, $T$. The number of water molecules $N_{w}$  is fixed. The relevant thermodynamic potential to analyze, is then the Helmholtz energy. 

The Helmholtz energy difference can be determined for each step in \eqr{eq/Intro/02} given the  temperature, volume, and numbers of molecules: $(T, V, N_{w}, N_{\cheq}, N_{\coeq})$.  In the simulations, we keep the temperature, volume and number of water molecules fixed, and control the chemical potentials of the guest components. The Helmholtz energy becomes then a function of the number of guest components $N_{\cheq}$ and $N_{\coeq}$. \eqr{eq/Intro/03a} can be specified as
\begin{equation}\label{eq/Intro/03as}
\Delta_r F(N_{\coeq}, N_{\cheq}) = \Delta_{f} F(N_{\coeq}, N_{\cheq}) + \Delta_{w} F + \Delta_{h} F(N_{\coeq}, N_{\cheq})
\end{equation}

We have earlier calculated the Helmholtz $\Delta_h F $  of step (2c), and are now interested in step \eqref{eq/Intro/02a}.  The Helmholtz energy of liquid water with dissolved gases  is equal to \cite{ll5}
\begin{equation}\label{eq/Energy/01}
F = -p\,V + \mu_{w}\,N_{w} + \mu_{\cheq}\,N_{\cheq} + \mu_{\coeq}\,N_{\coeq}
\end{equation}
where $p$ is the pressure in the fluid, and $\mu_{w}$, $\mu_{\cheq}$ and  $\mu_{\coeq}$ are the chemical potentials of water, methane and carbon dioxide, respectively. The number of water molecules is fixed to the number used to construct the hydrate phase.  For step \eqref{eq/Intro/02a}, we have 
\begin{equation}\label{eq/Energy/02}
-\Delta_{f}F = F(N_{\cheq}, N_{\coeq}) - F(0, 0)
\end{equation}
Following the same procedure as in \cite{Wierzchowski2007}, see \cite{Glavatskiy2012}, we obtain
\begin{equation}\label{eq/Energy/06}
-\Delta_{f} F = \mu_{\cheq}^*\,N_{\cheq} + \mu_{\coeq}^*\,N_{\coeq}  - \int_{-\infty}^{\mu_{\cheq}^*}N_{\cheq}\,d\mu_{\cheq} - \int_{-\infty}^{\mu_{\coeq}^*}N_{\coeq}\,d\mu_{\coeq}
\end{equation}
where $\mu_{\cheq}^*$ and $\mu_{\coeq}^*$ are chemical potentials specified by a reference phase. 

We shall here vary the chemical potentials $\mu_{\cheq}$ and $\mu_{\coeq}$ of the guest components, in order to vary the number of the guest molecules in the fluid $N_{\cheq}(\mu_{\cheq}, \mu_{\coeq})$ and $N_{\coeq}(\mu_{\cheq}, \mu_{\coeq})$. 
The computational scheme we use includes a reference gas phase, which is in equilibrium with the guest gas inside the simulation box. The pressure of this reference gas phase is determined by the chemical potentials of the guest components. We consider the system at a hydrostatic pressure $p$ equal to the pressure of the reference gas phase. 

The Gibbs energy difference can be determined for each step in \eqr{eq/Intro/02} from the corresponding Helmholtz energies via the Legendre transformation: 

\begin{equation}\label{eq/Energy/07}
\begin{array}{rl}
\Delta_{f} G (p, y_{\coeq}) &= \Delta_{f} F(N_{\coeq}(p, y_{\coeq}), N_{\cheq}(p, y_{\coeq})) + p \Delta V_f(p, y_{\coeq})
\\
\Delta_{h} G (p, y_{\coeq}) &= \Delta_{h} F(N_{\coeq}(p, y_{\coeq}), N_{\cheq}(p, y_{\coeq})) + p \Delta V_h(p, y_{\coeq})
\end{array}
\end{equation}
The notation $N_{\coeq}(p, y_{\coeq})$ and $N_{\cheq}(p, y_{\coeq})$ is used to indicate, that the number of the guest molecules in the fluid or hydrate phases are now functions of gas pressure, $p$, and the mole fraction $y_{\coeq}$ of \co in the reference gas phase (the mole fraction of \ch is $y_{\cheq} = 1-y_{\coeq}$). Furthermore, $\Delta V_f$ and $\Delta V_h$ are changes in unit cell volumes relative to the volume that defines $F$, when we apply  a particular pressure $p$, with mole fraction $y_{\coeq}$ of the reference gas phase. 

The functions $V(p, y_{\coeq})$ are given by equations of state. For the hydrate, we have the standard Birch-Murnagham equation of state \cite{Ning2012, Birch1947}. 
\begin{equation}\label{eq/Energy/08}
p(V) = \frac{3}{2}\,K\left[\epsilon^{7/3} - \epsilon^{5/3}\right]
\end{equation}
where $K$ is the isothermal bulk modulus, $\epsilon \equiv V_0/V$ is the compression factor and $V_0$ is the hydrate volume at zero pressure. For the fluid the Peng-Robinson EOS is used \cite{PengRobinson1976}.

The Gibbs  energy change of step \eqref{eq/Intro/02b} can be determined analogously to \eqr{eq/Energy/07}. In this case the reference gas pressure is zero, and the Gibbs energy change becomes equal to the Helmholtz energy change, leading to

\begin{equation}\label{eq/Energy/10}
\Delta_{w} G = G_{\left[ N_{\waeq}\right]_{\mathrm{hydrate}}} - G_{\left[N_{\waeq}\right]_{\mathrm{fluid}}}= F_{\left[ N_{\waeq}\right]_{\mathrm{hydrate}}} - F_{\left[N_{\waeq}\right]_{\mathrm{fluid}}}= \Delta_{w} F  
\end{equation}
where $F_{\left[ N_{\waeq}\right]_{\mathrm{hydrate}}}$ is the Helmholtz  energy of zero-occupancy hydrate and $F_{\left[N_{\waeq}\right]_{\mathrm{fluid}}}$ is the Helmholtz  energy of the pure water, and corrspondingly for $G$. The Gibbs energy of the hydrate was calculated by Wierzchowski and Monson \cite{Wierzchowski2007}. The Gibbs energy of the fluid can be obtained from the Peng-Robinson equation of state (PR-EOS) with the ideal gas reference state \cite{Glavatskiy2008}:
\begin{equation}\label{eq/Energy/09}
F_{\left[N_{\waeq}\right]_{\mathrm{fluid}}} = -RT\,\ln\left[ \frac{eV}{N\Lambda^3} \left(1-\frac{BN}{V}\right) \right] - \frac{1}{2\sqrt{2}}\frac{A}{B} \ln\left[ 1 + \frac{2\sqrt{2}\,BN}{V+BN(1-\sqrt{2})} \right]
\end{equation}
Here, $A$ and $B$ are the PR-EOS constants for water. \eqr{eq/Intro/03a} can now be specified as 
\begin{equation}\label{eq/Intro/03bs}
\Delta_r G(p, y_{\coeq}) = \Delta_{f} G(p, y_{\coeq}) + \Delta_{w} G + \Delta_{h} G(p, y_{\coeq}) 
\end{equation}
The notation for the differences in Gibbs energy follows the one used for the Helmholtz energy. 

\section{Simulation details}\label{sec/Simulations}
%


We performed GCMC ($\mu VT$) simulations of \ch, \co and \ho in a box of fixed volume.  The reference gas phase contained and the guest molecules only.  The volume of the box was equal to the volume of a 2x2x2 unit cell of sI hydrate, which has 64 cages. The cubic unit cell has a lattice parameter of 12.03 {\AA}\cite{McMullan1965a}, so the box size is 24.06 {\AA}. The system becomes equilibrated rather quickly, typically after 300-500 cycles. The number of cycles, during which the statistics for averaging is accumulated, was equal to 5000. The number of MC moves per cycle was equal to the number of particles of each component in the system, minimum 20.


All simulations were performed at 278 K. Pressure and fugacities varied in the range 10$^4$ Pa to 10$^9$ Pa. The value of the bulk modulus $K$ of a hydrate was 9 GPa \cite{Ning2012}. 

During the simulation, the guest molecules in the box are being equilibrated with the molecules in the reference phase. 
The chemical potential of the $i$-th component directly follows from its fugacity $f_{i}$ \cite{FrenkelSmit}: $\mu_{i}(T, f_{i}) = \mu_{i}^{0}(T) + k_{B}\,T\,\ln(f_{i}/f^{0})$, where $\mu_{i}^{0}$ and $f^{0}$ are ideal gas reference values, which do not affect the calculated values of $\Delta_{r} F$ and $\Delta_{r} G$. 
For pure components, fugacities were found from pressures using the Peng-Robinson equation of state. This procedure describes the pressure of the gas phase well, far from the critical point and was found to be rather accurate for the description of experimental data \cite{PengRobinson1976}. Furthermore, it corresponds well to the interaction potentials used \cite{GarciaPerez2007} for \co and \chh. The values of the critical temperature, critical pressure and the accentric factor used in PR-EOS for \co and \ch are\cite{GarciaPerez2007} respectively, 304.1282 K and 190.564 K, 7377300.0 Pa and 4599200.0 Pa, 0.22394 and 0.01142. The corresponding values for water used in the PR-EOS are \cite{GarciaPerez2007} 647.14 K, 22064000.0 Pa and 0.3443. 

For mixtures, we applied the Lewis-Randall rule \cite{moranshapiro} to convert pressures into fugacities, from their pure phase values to mixture values. The procedure is not exact as the mixture of methane and carbon dioxide behaves as a regular solution \cite{Barry1982}. However, it is common to approximate the mixture as an ideal solution \cite{Arai1971} because the excess enthalpy is small.


We used the TIP5PEw model of water\cite{TIP5PEW}. The description of the guest molecules was taken from the TraPPE force field \cite{TRAPPE}. We used the LJ interaction potential, truncated and shifted to zero at 12 {\AA}. The parameters of the potential are specified in \tblr{tbl/LJ} and the parameters of atoms are specified in \tblr{tbl/atoms}. Ewald summation was used with \cite{Calero2004} a relative precision $10^{-6}$ and alpha convergence parameter $0.265058$ {\AA}$^{-1}$. 


The value of $G_{\left[ N_{\waeq}\right]_{\mathrm{clathrate}}} $ referred to Gibbs energy of an ideal gas, was given by \cite{Wierzchowski2007} as -38.67 kJ/mol water for a 2x2x2 lattice at 241.5 K and 102.9 bar. This pressure leads, according to \eqr{eq/Energy/07}, to the value $F_{\left[ N_{\waeq}\right]_{\mathrm{clathrate}}}$ equal to -124.96 kJ/mol for 2x2x2 lattice or -15.62 kJ/(mol of unit cell) (ideal gas reference). We assumed that the temperature variation over 35 K is small (see discussion below).  \eqr{eq/Energy/09} leads to $F_{\left[ N_{\waeq}\right]_{\mathrm{fluid}}}$ equal to -40.79 kJ/(mol of unit cell) using the same reference state. The energy difference for step (2b)  becomes, according to \eqr{eq/Energy/10}, $\Delta_w F = \Delta_{w} G = 25.17$ kJ/mol.


Semi-grand canonical Monte Carlo (MC) simulations were performed \cite{Glavatskiy2012} of the liquid water mixture, where we specified the number of water molecules $N_{w}$ to be equal to the number in the constructed hydrate. In the 2x2x2 unit cell there are 368 water molecules which are initially placed randomly in the box (one unit cell has 46 water molecules). Furthermore, we specified the fugacities of the guest components $f_{\coeq}$ and $f_{\cheq}$  and calculated the loading of the guest components $N_{\coeq}$ and $N_{\cheq}$. The Helmholtz energy of the fluid mixture was calculated from the absorption isotherm according to \eqr{eq/Energy/06} and the Gibbs  energy of step (2a) was calculated from \eqr{eq/Energy/07}. In our GCMC simulations the guest molecules were allowed to change their position and orientation. They were also subjected to Regrow, Swap, and Identity change \cite{Martin1997} MC moves. Furthermore, unlike in the simulations of the hydrate \cite{Glavatskiy2012}, the positions of the water molecules were not fixed: they were also allowed to change their position and orientation. 

\section{Results and discussion}\label{sec/Results}

In the presentation of the results for the water fluid phase, we shall use the  "`cell loading"' as a composition measure. The "`cell loading"' means the number of molecules per unit cell, [m./u.c.], equal to the number of gas molecules dissolved per 46 water molecules. This makes possible a direct comparison with results \cite{Glavatskiy2012} for guest molecules in hydrate lattices. The hydrate lattice has a maximum number of 8 positions for guest molecules.

\begin{figure}[ht!]
\centering
\subfigure[ ] %
{\includegraphics[scale=\scaleprofile]{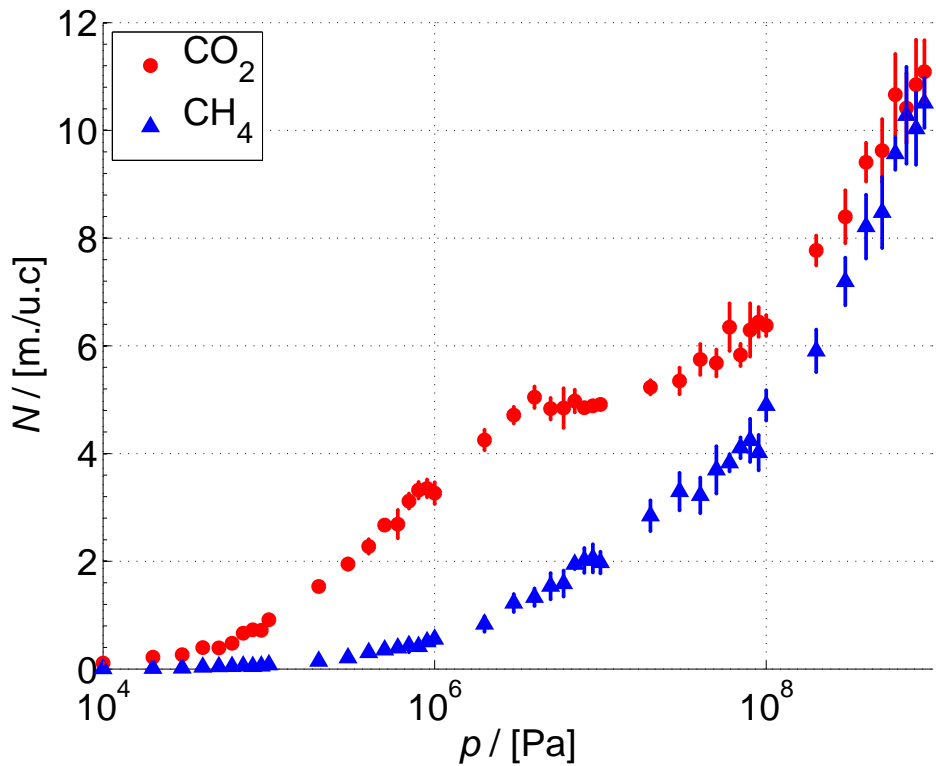}\label{fig/Np}} %
\subfigure[ ] %
{\includegraphics[scale=\scaleprofile]{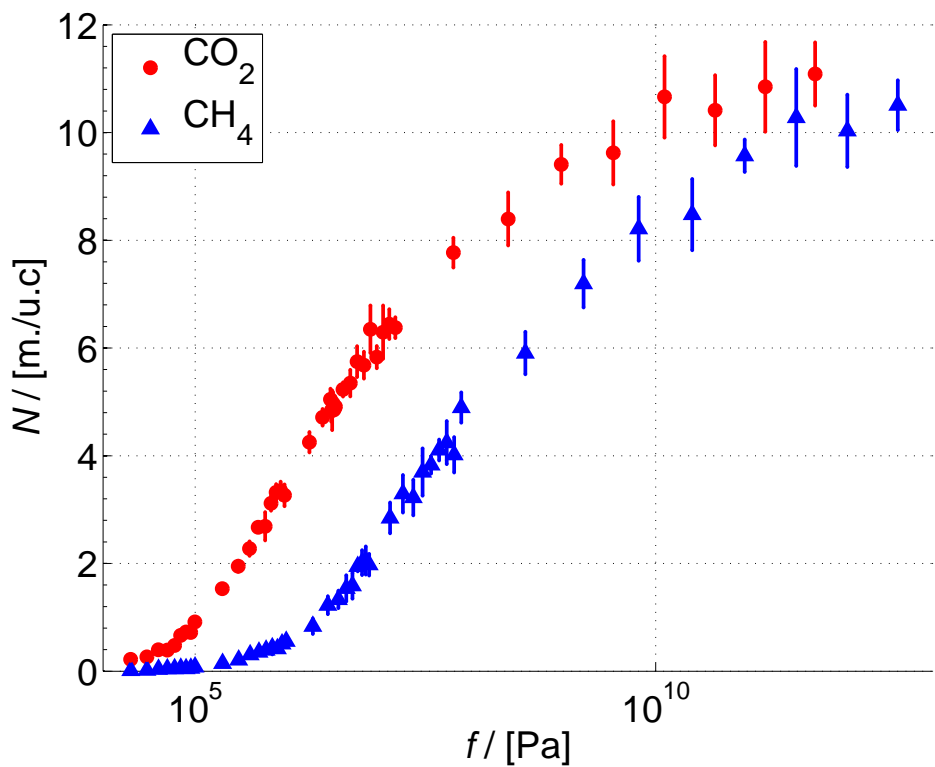}\label{fig/Nf}} %
\caption{Number of guest molecules per fluid unit cell in \co+water and \ch+water as a function of the reference gas pressure (a) and fugacity (b) as computed by GCMC simulations at $T = 278$ K for step \eqref{eq/Intro/02a}.}\label{fig/loading}
\end{figure}

The cell loading, defined in this manner, was plotted in \figr{fig/loading} as a function of pressure and fugacity for \co and \ch pure gas phase. The loading profile at the constant temperature may be regarded as the fluid isotherm. We note that, unlike for the hydrate phase, more than 8 molecules can fit in the same volume when water is in the fluid phase. This is expected when the water molecules can move freely within the box, without forming cages.

\begin{figure}[ht]
\centering
\includegraphics[scale=\scaleprofile]{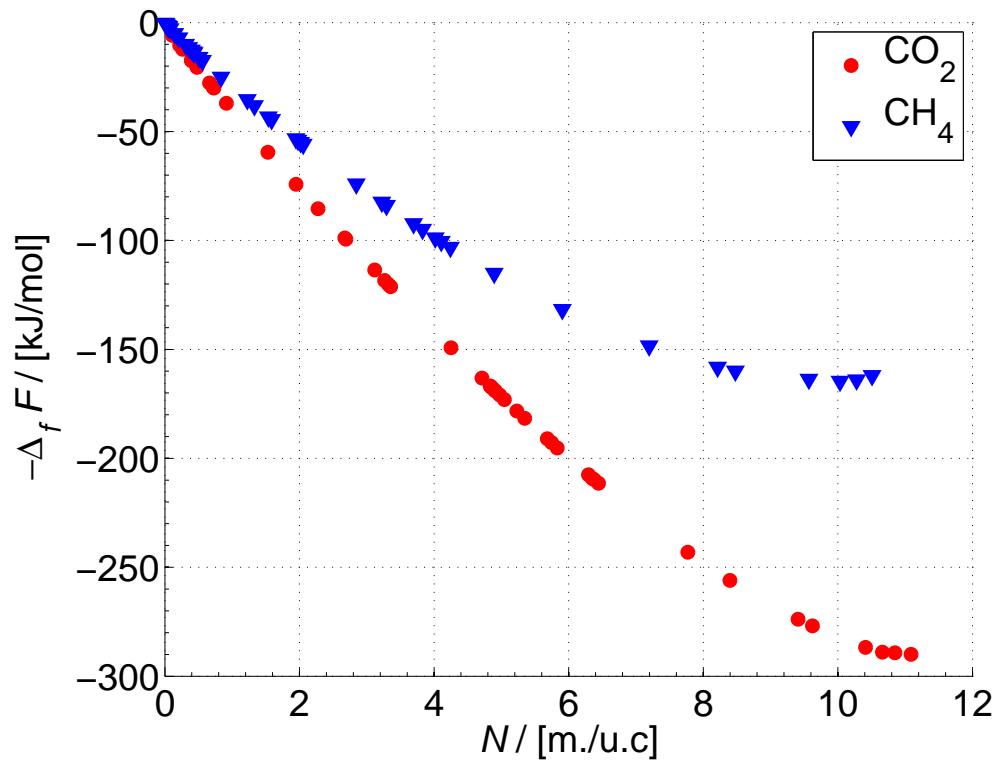}
\caption{Helmholtz energy difference of step \eqref{eq/Intro/02a} for \co+water and \ch+water at 278 K and constant volume, as a function of the number of guest molecules per fluid unit cell.}\label{fig/FN_f}
\end{figure}

By integrating these isotherms using \eqr{eq/Energy/06}, we obtained $\Delta_{f}F$. \figr{fig/FN_f} shows the Helmholtz energy per 46 water molecules (\textit{i.e.} per mole of unit cell), for fluids with \co+\ho or \ch+\ho, as a function of guest loading. The Helmholtz energy difference decreases with the number of molecules adsorbed in both cases. One can also see that the Helmholtz energy difference for the \co+\ho fluid is significantly lower than that of the \ch+\ho fluid, over the whole range of compositions, meaning that \co+\ho is more stable, when we compare states with the same volume. Also, the difference in the curves for hydrate and fluid phase  is smaller for \co than for \ch.  Both properties can be attributed to the polar character of \co and is in agreement with experimental findings at 277 K and 8.2 MPa \cite{Graue}. 
 
\begin{figure}[ht]
\centering
\subfigure[] %
{\includegraphics[scale=0.7]{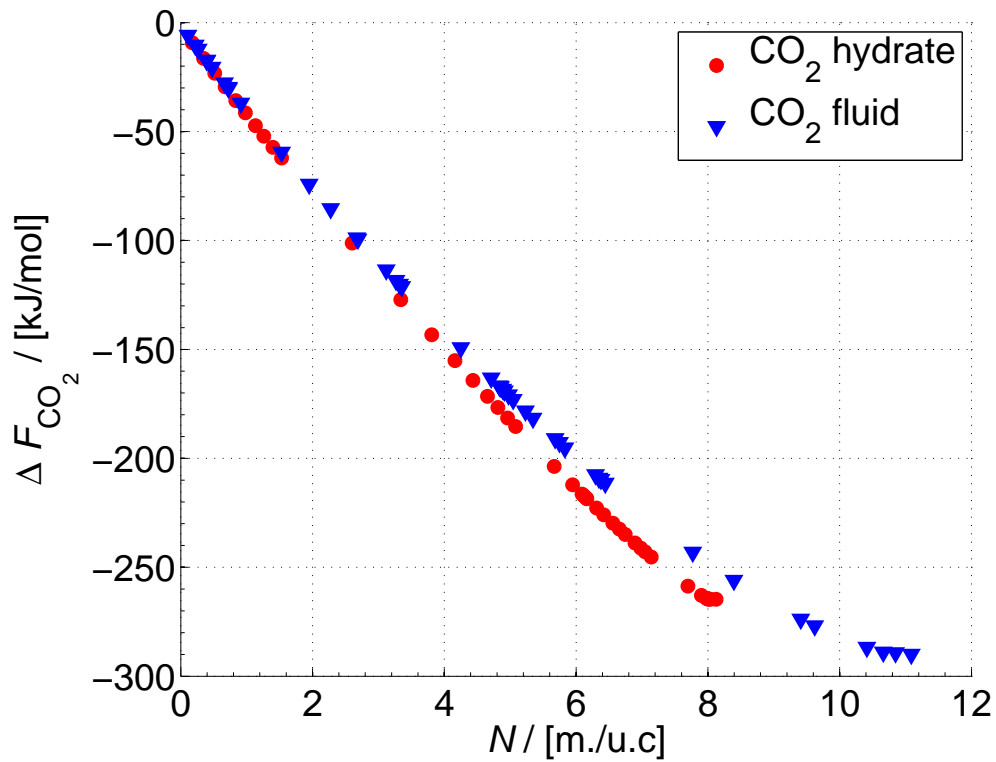}\label{fig/FN_CO2}}
\subfigure[] %
{\includegraphics[scale=0.7]{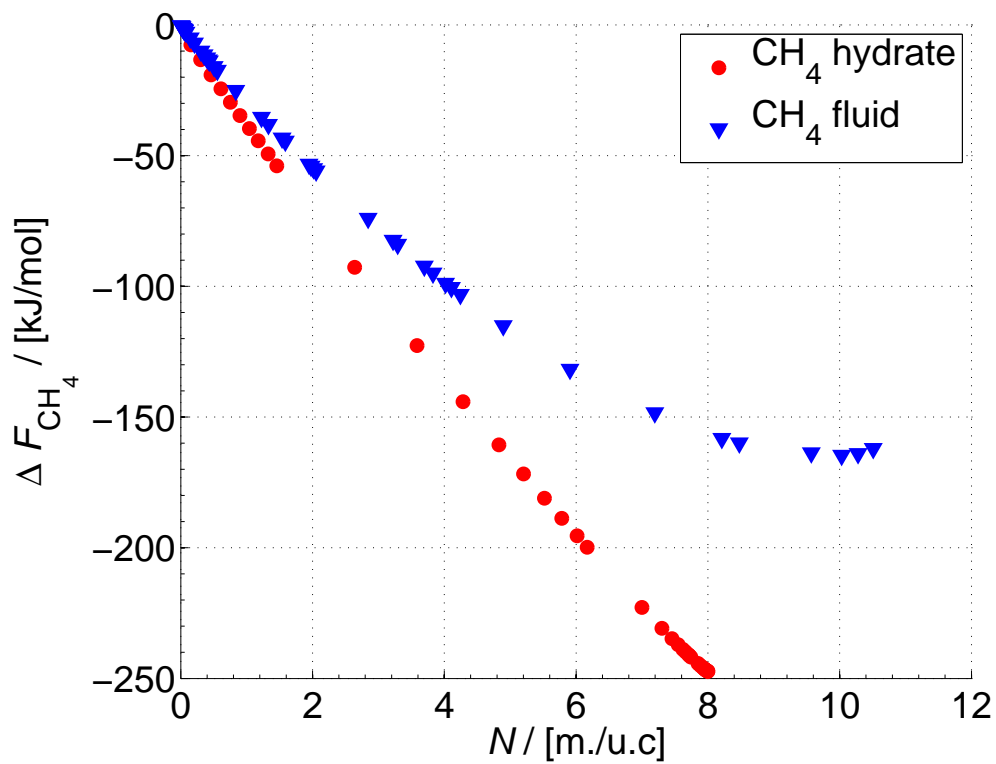}\label{fig/FN_CH4}}%
\caption{Helmholtz energy difference of steps \eqref{eq/Intro/02a} (fluid) and \eqref{eq/Intro/02c} (hydrate) for (a) \co+water  (b) \ch+water at 278 K and constant volume as a function of the number of guest molecules per unit cell.}\label{fig/FN}
\end{figure}

By comparing the Helmholtz energy of the fluid and hydrate phases for each of the guest components (see \figr{fig/FN}), we see that the value of the Helmholtz energy for the hydrate phase is always lower than the value of the fluid phase. This theoretical results is supported by observations \cite{Graue} that hydrates form at 278 K and  between 10$^4$-10$^9$ Pa. 

\begin{figure}[ht]
\centering
\subfigure[] %
{\includegraphics[scale=0.6]{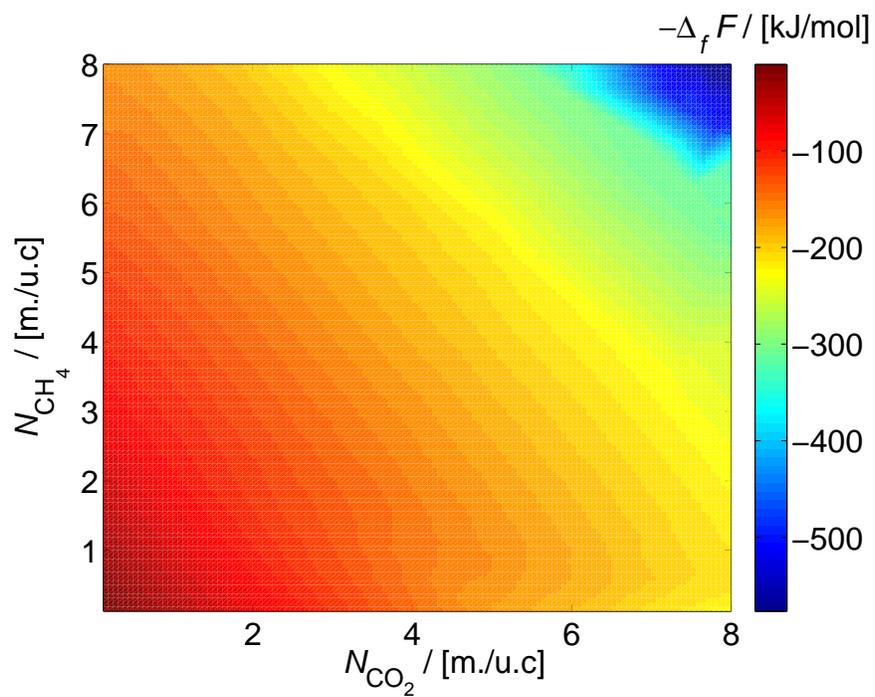}\label{fig/FNN_f_CO2+CH4}}
\subfigure[] %
{\includegraphics[scale=0.6]{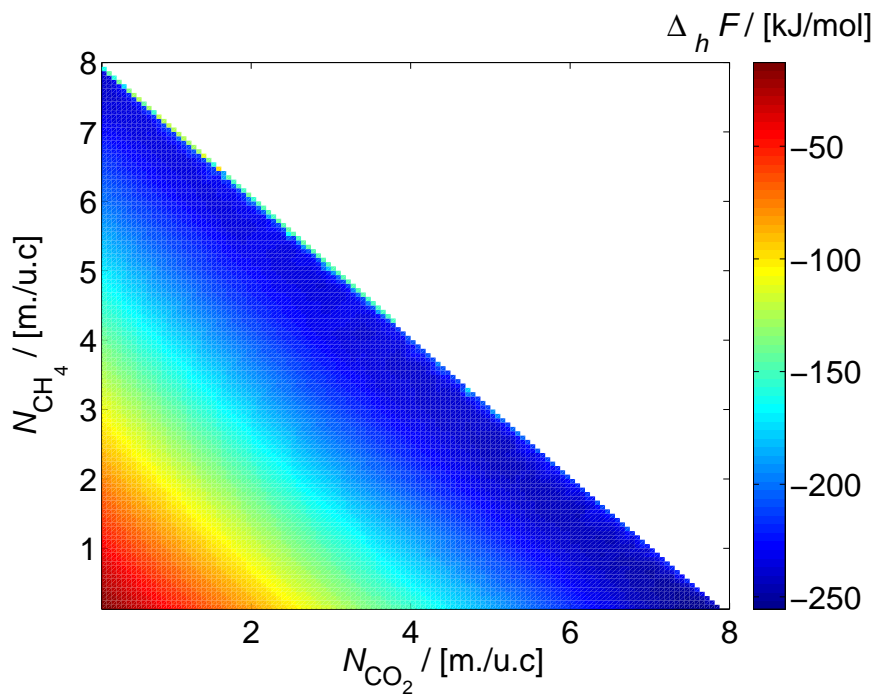}\label{fig/FNN_h_CO2+CH4}}%
\caption{Helmholtz energy difference of steps \eqref{eq/Intro/02a} (a) and \eqref{eq/Intro/02c} (b) as computed by \eqr{eq/Energy/06} for the \co+\ch gas mixture at 278 K and constant volume as a function of the loading the components.}\label{fig/FNN_fh_CO2+CH4}
\end{figure}

We present the Helmholtz energy of the mixture \co+\ch in the fluid and in the hydrate in terms of loading of each component in \figr{fig/FNN_fh_CO2+CH4}. The figure shows results for the fluid and hydrate phases. Because the total number of molecules per unit cell is less than or equal to 8 in the hydrate phase, the top right corner of  \figr{fig/FNN_h_CO2+CH4} is empty. The points on the diagonal correspond to the total loading slightly higher than 8. This leads to the value of the Helmholtz energy for these points much higher than the one for the maximum allowed loading of 8 molecules (this is also the case for the total Helmholtz energy presented in the next paragraph). There is no such restriction for fluid phase, as is seen in \figr{fig/FNN_f_CO2+CH4}. Increasing the loading both in the hydrate and in the fluid decreases the Helmholtz energy difference, \textit{i.e.} both phases become relatively more stable. These data are used to calculate the total Helmholtz energy difference for the whole reaction \eqref{eq/Intro/01}. 

\begin{figure}[ht]
\centering
{\includegraphics[scale=0.7]{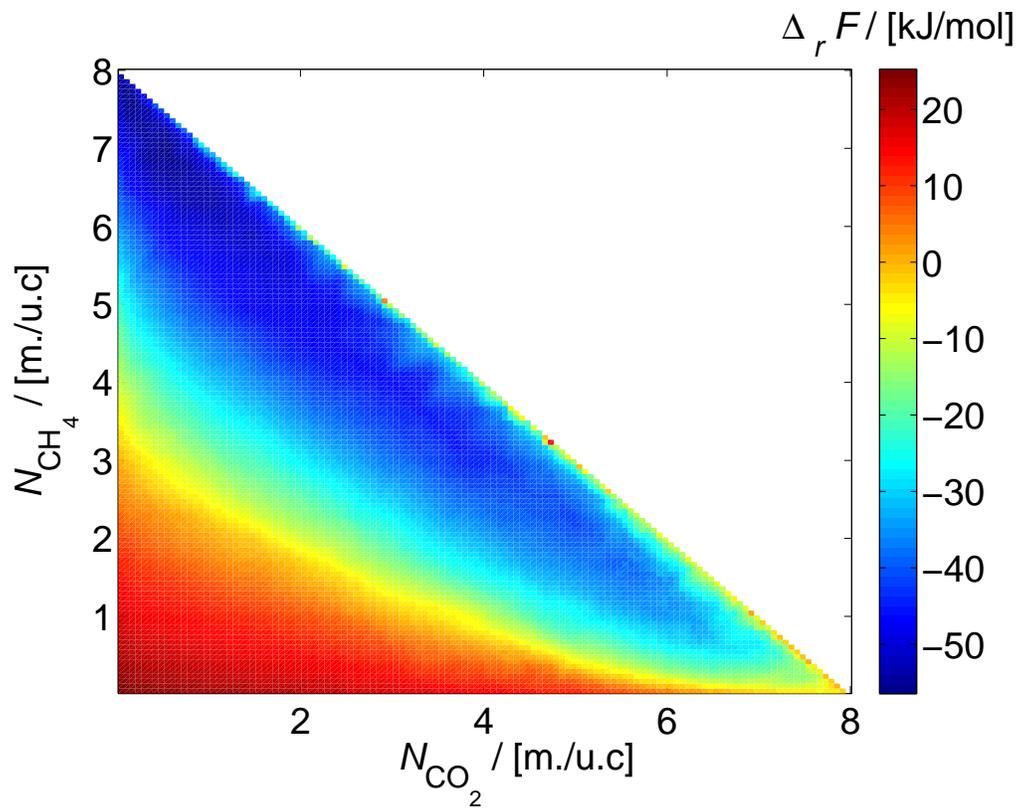}%
\caption{Helmholtz energy difference of the formation reaction \eqref{eq/Intro/01} for the \co+\ch gas mixture at 278 K as a function of hydrate loading of the components.} \label{fig/FNN_CO2+CH4}}
\end{figure}

The Helmholtz energy for the whole reaction \eqref{eq/Intro/01} is plotted in \figr{fig/FNN_CO2+CH4} as a function of the hydrate loading. The lowest value is reached at high loading, consistent with the findings presented in \figr{fig/loading}. This suggests that a fully occupied hydrate can form at a particular volume and temperature. The compositions for which $\Delta_r F = 0$, are equilibrium compositions for gases in equilibrium with the volume-constrained hydrate.  These compositions, which can be found by experiment and used to test our calculations, correspond to a total loading of 3-6 molecules per unit cell (the light blue region in the figure).  Stevens et al. \cite{Stevens} found that hydrates with a mixture of methane and carbon dioxide  formed at 277 K and 8.3 MPa. 

\begin{figure}[ht]
\centering
\subfigure[] %
{\includegraphics[scale=0.6]{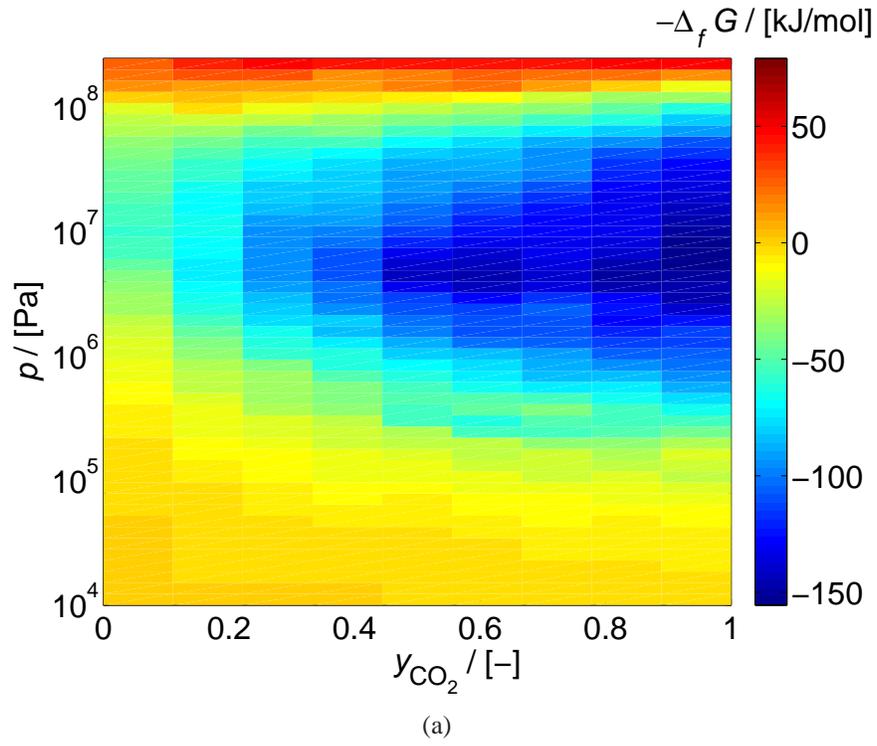}\label{fig/GPY_f_CO2+CH4}}
\subfigure[] %
{\includegraphics[scale=0.6]{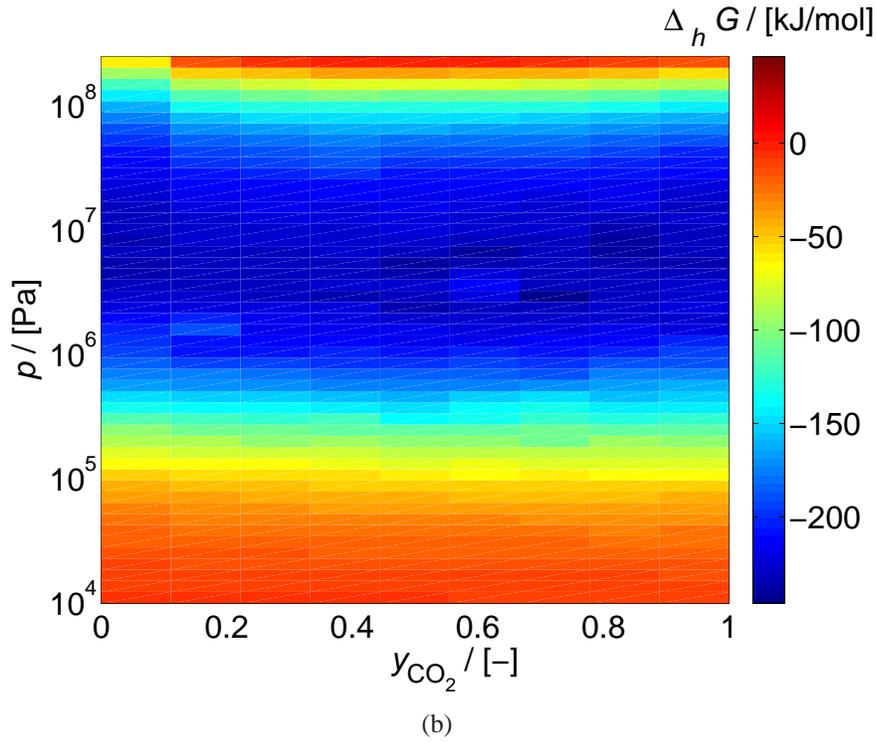}\label{fig/GPY_h_CO2+CH4}}%
\caption{Gibbs  energy difference of steps \eqref{eq/Intro/02a} (a) and \eqref{eq/Intro/02c} (b) as computed by \eqr{eq/Energy/07} for the \co+\ch gas mixture at 278 K as a function of the reference gas pressure and mole fraction of \co in the reference gas phase.} \label{fig/GPY_fh_CO2+CH4}
\end{figure}

The Gibbs  energy for reaction steps (2a) and (2c) are plotted in \figr{fig/GPY_fh_CO2+CH4}. To provide a link between operating conditions and hydrate stability questions, the results are represented by $p$-$y$ diagrams, where $p$ is the pressure of the reference gas corresponding to the components' chemical potentials,  and $y_{\coeq}$ is the mole fraction of \co in the reference gas phase. In the present calculations, gases were in equilibrum with the hydrate at the pressure of the reference phase. In a real situation, the equilibrium is established at a hydrostatic pressue. The fluid phase is more stable at higher pressures (above 10$^7$ Pa) than at lower pressures. Typical operating pressures are around 10$^7$ Pa.

\begin{figure}[ht]
\centering
{\includegraphics[scale=0.7]{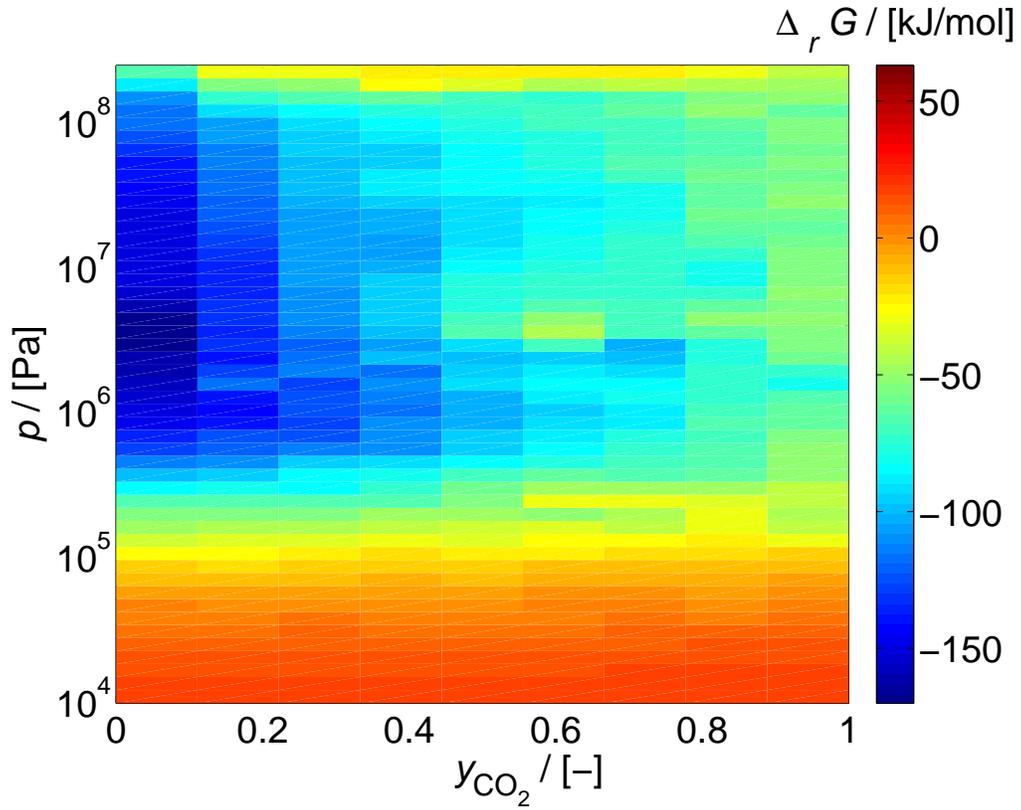}%
\caption{Gibbs energy difference of the formation reaction \eqref{eq/Intro/01} for the \co+\ch gas mixture at 278 K as a function of the reference gas pressure and mole fraction of \co in the reference gas phase.}\label{fig/GPY_CO2+CH4}}
\end{figure}

Finally, the Gibbs  energy change of the total reaction \eqref{eq/Intro/01} is plotted in \figr{fig/GPY_CO2+CH4}. This is the quantity which decides on the direction of the formation of hydrates at constant temperature and pressure. When $\Delta_{r}G$ is negative, the direction of the reaction is from the left to the right. When $\Delta_{r}G$ is positive, the reaction goes from the right to the left. We see from \figr{fig/GPY_CO2+CH4} that  $\Delta_{r}G<0$ when the pressure lies between 10$^5$ and 10$^7$ Pa over the whole range of mole fractions of \coo. This is where formation of hydrate is favorable at constant $p$ and $T$. At low pressures $\Delta_{r}G$ is positive and the hydrate will dissociate at any mole fraction of \coo. When $\Delta_{r}G = 0$, we have equilibrium for the formation reaction \eqref{eq/Intro/01}. These values should be observable in experiments. 

In ref. \cite{Glavatskiy2012} we speculated about a possible thermodynamic path to convert the methane hydrate to the carbon dioxide hydrate, based on data for step (2c) only. From the $p$-$y$ diagram of the Helmholtz energy for the hydrate with a mixture of gases  \cite{Glavatskiy2012}, we concluded that there exists a region of negative energies across which it might be possible to go from pure methane hydrate to pure carbon dioxide hydrate. According to that path one should first decrease the pressure in a (high-pressure) deposit of methane hydrate to a value in the range of 10$^6$-10$^7$ Pa. This is the pressure range where all small clathrate cages are empty, and methane begins to de-occupy the large cages. The large cages can then be filled with the carbon dioxide molecules without extra energy. Therefore, it was suggested to gradually increase the fraction of \co, keeping the pressure fixed. (Increasing the pressure would increase the energy required to put the \co molecule in the small cage, while decreasing the pressure would destabilize the clathrate.) After all of the methane had been exchanged with carbon dioxide, we proposed to increase the pressure to fill the entire clathrate with carbon dioxide. With \figsr{fig/FNN_CO2+CH4}{fig/GPY_CO2+CH4} we can now test this hypothesis for the whole reaction for the total reaction (1) under various conditions.


The Gibbs energy change of the total reaction, shown in \figr{fig/GPY_CO2+CH4},  is zero for pressures around 10$^5$ Pa. It means that below this pressure the reaction \eqref{eq/Intro/01} will go from the right to the left, i.e. a stable hydrate is not possible. In contrast, for the pressures larger than 10$^5$ Pa, the formation reaction will progress from the left to the right, which means that a stable hydrate can exist. If we start with fully occupied methane hydrate, we are in the top left corner. The pressure can be reduced to 10$^6$ Pa, keeping a methane hydrate stable. The small cages will be first emptied, and the large cages will also begin to be de-occupied as we decrease the pressure. If now one increases the mole fraction of \co in the reference gas phase, $\Delta_{r} G$ will be almost as low at all mole fractions of \co as at zero mole fraction of \co. Increasing the \co mole fraction at another value of the pressure may still keep the hydrate stable ($\Delta_{r} G < 0$), but will decrease its relative stability compared to the pressure of 10$^6$ Pa. In particular, increasing the \co mole fraction at the highest pressure, when the hydrate is fully occupied, may lead to the state at which the hydrate will become unstable. When all methane is replaced with carbon dioxide in large cages through this mechanism, one can start increasing the pressure. This will lead to \co occupation of the small cages as well, until the hydrate becomes fully occupied. The path described is therefore feasible. 

In \figr{fig/FNN_CO2+CH4} the equilibrium line with $\Delta_r F = 0$ corresponds to a total loading of the hydrate of 3 molecules per unit cell for pure \ch, and 6 molecules per unit cell with pure \co. The last value is the same as the number of large cages, which indicates that this is an upper bound for equilibrium with \co. To have only 3 cages filled with \ch, means that the remaining cages can be filled with \co. We know that the larger cages are filled first. During this process, reaction \eqref{eq/Intro/01} will be in equilibrium, while the total composition will change. The picture in terms of loading reveals the mechanism according to which the exchange could happen. 

The value of $\Delta_{w} G$ was taken from the literature \cite{Wierzchowski2007} at slightly different conditions than ours. A change in its value will shift $\Delta_{r} G$ or $\Delta_{r} F$ by a constant. We see from \figsr{fig/FNN_CO2+CH4}{fig/GPY_CO2+CH4} that shift up to 50 kJ/mol will not alter the conclusions above. Taking into account that the value of $\Delta_{w} G$ is 25.17 kJ/mol, we can conclude that it is safe to use this value.

To summarize; we have found that it seems feasible, based on thermodynamic arguments, that \ch can be processed while \co can be stored, for process operating conditions where $V, T$ or $p, T$ are constant. We have determined regions where the exchange is spontaneous or can be perturbed by shifting equilibria. This can give theoretical support to field test now being carried our in Alaska \cite{alaskalink}. 

The equilibrium relations, which derive from the results, can be tested in the laboratory, to establish the accuracy of the modeling.  Experimental tests should expose the hydrate at equilibrium to a varying external gas pressure. By decreasing the pressure of the gas mixture, one may bring a pure \ch hydrate first to a step where all small and a few large cages are empty. At this moment, one may start to add \co until all large cages are filled, gradually forcing \ch out. When no \ch is left in the hydrate, one may increase the pressure back to the starting one, to obtain a pure \co hydrate. The actual values of the operating pressures may differ from the ones predicted here due to conditions not used here, but imposed by the real system, for instance variations in rock porosity and surface conditions.  
 
From the kinetic point of view it may, however, be complicated to perform an exchange of one type of molecules with another in a hydrate. Unlike in zeolites, a hydrate does not have empty channels for molecules to traverse. One way to realize the exchange in hydrates, could be by first distorting the  clathrate structure \cite{Peters2008}.

\section{Conclusions}\label{sec/Conclusions}

We have presented results for thermodynamic stability of sI hydrate formation with \co and \ch. The formation was considered as three consecutive steps, where the last one was discussed earlier \cite{Glavatskiy2012}.  We performed missing grand-canonical Monte Carlo simulations, and obtained Helmholtz and Gibbs energy criteria for hydrate formation.  The results suggest that there exist at constant $V,T$, and at constant $p,T$, a thermodynamic path which favors the formation of \co hydrate from the \ch hydrate. We expect that the route proposed can be tested experimentally. This would give valuable information for the processing of gas hydrates from porous rocks or bulk layers on the ocean floor.

\section{Acknowledgment}
We are grateful to the Statoil-VISTA grant \#6343.

\clearpage
\clearpage
\clearpage
\clearpage
\clearpage
\clearpage
\clearpage
\clearpage
\begin{longtable}{l@{\qquad}l@{\qquad}l}%
\caption{Parameters of the LJ interaction potential between C and O atoms in $\cheq$, $\coeq$ and $\waeq$ molecules \cite{GarciaPerez2007}.} \label{tbl/LJ}\\%
\hline
\hline
atomic pairs & $ K\epsilon$/($k_B$) & $\sigma$/{\AA}  \\
\hline %
 C$_\cheq$ - C$_\cheq$ & 158.50000000 & 3.72000000 \\
 C$_\cheq$ - C$_\coeq$ & 68.87985475 & 3.23237650 \\
 C$_\cheq$ - O$_\coeq$ & 116.52842617 & 3.36866350 \\
 C$_\cheq$ - O$_\waeq$ & 119.11459189 & 3.40850000 \\
 C$_\coeq$ - C$_\coeq$ & 28.12900000 & 2.76000000 \\
 C$_\coeq$ - O$_\coeq$ & 47.59000000 & 2.89000000 \\
 C$_\coeq$ - O$_\waeq$ & 51.76401128 & 2.92087650 \\
 O$_\coeq$ - O$_\coeq$ & 80.50700000 & 3.03300000 \\
 O$_\coeq$ - O$_\waeq$ & 87.57246641 & 3.05716350 \\
 O$_\waeq$ - O$_\waeq$ & 89.51600000 & 3.09700000 \\
\hline %
\hline
\end{longtable} %

\clearpage
\begin{longtable}{l@{\qquad}l@{\qquad}l@{\qquad}l}%
\caption{Mass $m$, charge $q$ and atom size $r$ in $\cheq$, $\coeq$ and $\waeq$ molecules \cite{GarciaPerez2007}. The atoms $i$ and $j$ are considered as 'bonded' if the distance between the atoms is smaller than $0.56$~{\AA}~$\,+\,r_i\,+\,r_j$.} \label{tbl/atoms}\\%
\hline
\hline
atoms& $m$ / $1.6605402 \times 10^{-27}$ kg & $q$ / $1.60217733 \times 10^{-19}$ C  & $r$ / {\AA}  \\
\hline %
C$_\cheq$ &  16.04246 &   0.0  & 1.00\\
C$_\coeq$ & 12.0  & 0.6512  & 0.720 \\
O$_\coeq$ & 15.9994 &  -0.3256  & 0.68 \\
O$_\waeq$ & 15.9994 &    0.0    & 0.5  \\
H$_\waeq$ & 1.008   &   0.241  & 1.00\\
Dummy H$_\waeq$ & 0.0     &  -0.241  & 1.00\\
\hline %
\hline
\end{longtable} %

\clearpage

\bibliography{hydrate}

\clearpage
\begin{figure}[ht]
\centering
{\includegraphics[scale=0.3]{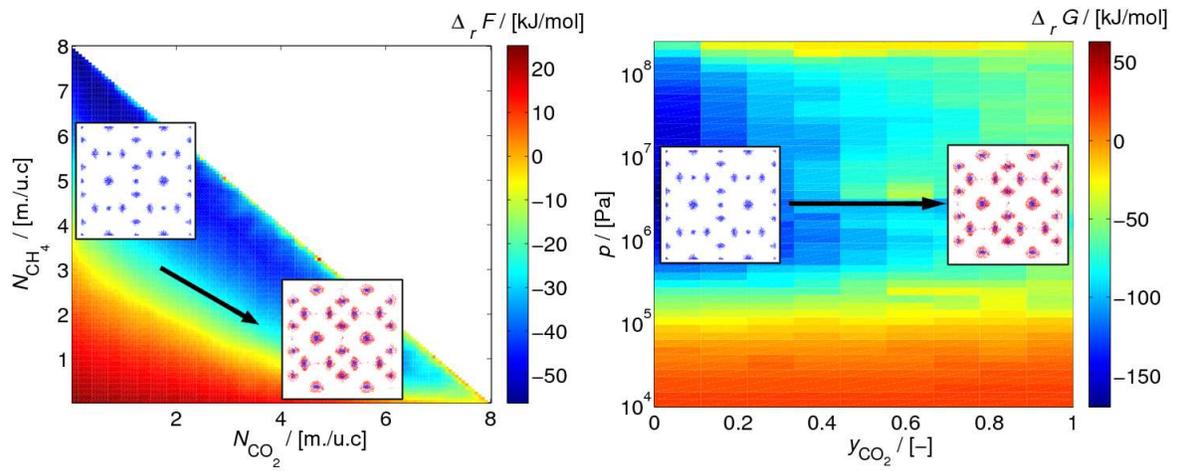}}
\caption*{TOC figure}
\end{figure}

\end{document}